# Nanosecond Reversal of Three-Terminal Spin Hall Effect Memories Sustained at Cryogenic Temperatures

Graham E. Rowlands,[1, a)] Minh-Hai Nguyen,[2] Sriharsha V. Aradhya,[2] Shengjie Shi[2], Colm A. Ryan,[1] Robert A. Buhrman,[2] and Thomas A. Ohki[1]

[1)]*Raytheon BBN Technologies, Cambridge, MA 02138*
[2)]*Cornell University, Ithaca, New York 14853, USA*

***Abstract:*** *We characterize the nanosecond pulse switching performance of the three-terminal magnetic tunnel junctions (MTJs), driven by the spin Hall effect (SHE) in the channel, at a cryogenic temperature of 3 K. The SHE-MTJ devices exhibit reasonable magnetic switching and reliable current switching by as short pulses as 1 ns of $<10^{12}$ A/m$^2$ magnitude, exceeding the expectation from conventional macrospin model. The pulse switching bit error rates reach below $10^{-6}$ for $\leq 10$ ns pulses. Similar performance is achieved with exponentially decaying pulses expected to be delivered to the SHE-MTJ device by a nanocryotron device in parallel configuration of a realistic memory cell structure. These results suggest the viability of the SHE-MTJ structure as a cryogenic memory element for exascale superconducting computing systems.*

After many years of steady progress, magnetoresistive random access memories (MRAMs) based on spin-transfer torque (STT) [1,2] switching are poised to complement and perhaps replace conventional DRAM architectures. One motivation for pursuing these novel architectures is their reduced static power consumption afforded by the intrinsic non-volatility of MRAM elements. Despite such a reduction, high-performance computing platforms will soon reach performance ceilings imposed by the power consumption of their underlying CMOS logic. As such, CMOS is being targeted for eventual replacement. This has caused a resurgent interest in Josephson junction (JJ) logic (an original competitor to CMOS), which could enable exascale computing at 1/100 the power of CMOS systems despite the power required for cooling the junctions below their superconducting transition temperature [3]. Such a system requires a fast cryogenic memory; preferably one that provides the same benefits offered by STT devices at room temperature. While notable progress has been made in creating memories that utilize both magnetic and superconducting effects [4], these devices are difficult to switch electrically and cannot presently leverage much of the extensive MRAM toolkit — synthetic antiferromagnets, pinning layers, etc. — given the adverse impacts of such structures on superconducting thin film circuits. We therefore restrict our attention to "normal" devices that operate by the same principles at cryogenic temperatures as they do at room temperature.

Barring any auxiliary phenomena, anti-damping (AD) STT switching relies on advantageous thermal fluctuations to create an initial torque in devices with colinear in-plane magnetized (IPM) magnetic layers, and thus will operate neither quickly nor reliably at cryogenic temperatures. Accordingly, most demonstrations of low-temperature STT switching have utilized non-collinear spin torque from a perpendicular polarizer [5,6] or a canted in-plane polarizer [7] to increase the initial STT (as has also been done at room temperature [8–10]). Recently, however, it has been demonstrated that three-terminal magnetic tunnel junctions based on the spin-Hall effect (SHE-MTJs) do not suffer from these limitations, most likely due to the non-uniform nature of the magnetic reversal [11] in combination with the beneficial in-plane Oersted field generated by the pulse current in the spin Hall channel [12]. Remarkably, it has been recently demonstrated that SHE-MTJs devices can be switched with sub-ns pulses and can exhibit write error rates (WERs) below $10^{-5}$ for pulses as short as 2 ns without fear of dielectric tunnel barrier breakdown [13]. This IPM approach offers considerable simplicity over other spin-torque device alternatives: precessional switching in multi-polarizer devices requires very precise timing inherent to precessional reversal dynamics [14,15], while fast SHE switching of a perpendicular magnetized (PM) free layer requires some manner of in-plane effective-field bias (from exchange pinning, external fields, or other mechanisms) to break the axial symmetry of the FL, as well as precise pulse timing in the short pulse regime to obtain deterministic reversal [6]. Here we demonstrate that IPM 3T-MTJs can continue to provide reliable and fast switching behavior down to $T = 3$ K without the need for involving these additional complexities. We also examine the potential compatibility of these devices with write pulses from superconducting nano-cryotron (nTron) drivers [16], finding an unexpected performance improvement.

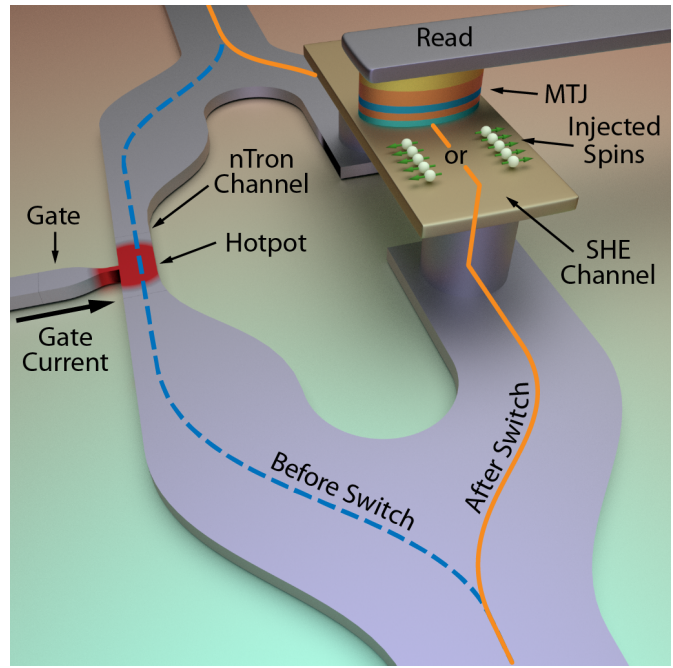

*Figure 1: Schematic of the proposed memory unit cell with an nTron select element and SHE-MTJ element. Current flows through the nTron channel but can be redirected through the spin Hall channel by the application of a gate current that creates a hotspot in the indicated area. Spins with either indicated orientation are created by reversing the polarity of the current through the unit cell.*

Our devices consists of an IPM MTJ, utilizing an MgO barrier, patterned on top of a an approximately 200 nm wide Pt(5)/Hf(0.7) spin-Hall channel (thicknesses in nm). The choice of Pt is motivated by the positive sign of its spin-Hall angle, $\theta_{SH}$, for which the in-plane Oersted field from the switching current in the spin Hall channel assists the torque from the SHE. The result is fast reversal with minimal signs of any incubation delay [12]. The role of the Hf space is to reduce the undesired interfacial enhancement of the magnetic damping of the free layer (FL) arising from the interaction of Pt with the ferromagnetic material, without causing a significant detrimental effect on the effective spin torque efficiency $\xi_{SH}$ [17]. The FL is composed of $Fe_{60}Co_{20}B_{20}$ (1.6), and, due to the interface with the MgO tunnel barrier, possesses an effective perpendicular magnetic anisotropy field of approximately 1 T. This anisotropy reduces the FL's effective demagnetization field, and therefore its required switching current, while allowing the FL to remain magnetized in plane at both $T = 300$ K and $T = 3$ K. The reference layer (RL) consists of a FeCoB/Ru/FeCoB synthetic antiferromagnet (SAF) strongly exchange pinned by IrMn in order to help minimize polarity dependent switching behavior. Full device details are given in our previous work [11].

A schematic of our proposed memory cell is shown in Fig. 1. The SHE-MTJs cannot be driven directly by JJ logic, which operates at the level of individual flux quanta. Instead we intend to utilize nTron devices as select elements that are triggered by gate pulses from peripheral JJ logic. The gate pulses nucleate a hot-spot in the nTron channel constriction, causing it to enter the resistive state. For this current work, however, we study isolated memory elements using a fast switching measurement setup and cryogenic environment described elsewhere [18]. All measurements reported in this paper were performed at 3 K.

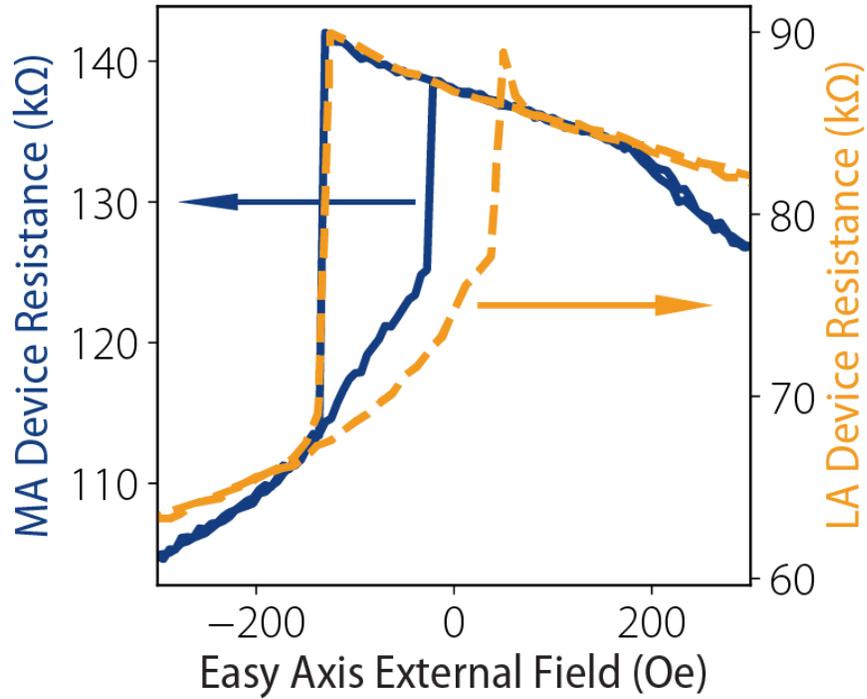

*Figure 2:* *Magnetic minor loops of a MA (left axis) and LA (right axis) SHE-MTJ device, showing a sharp AP→P switching but a gradual P→AP transition which can be attributed to the Néel coupling between the magnetic edge charges of the reference and free layer, and/or the complex multidomain micromagnetic reversal.*

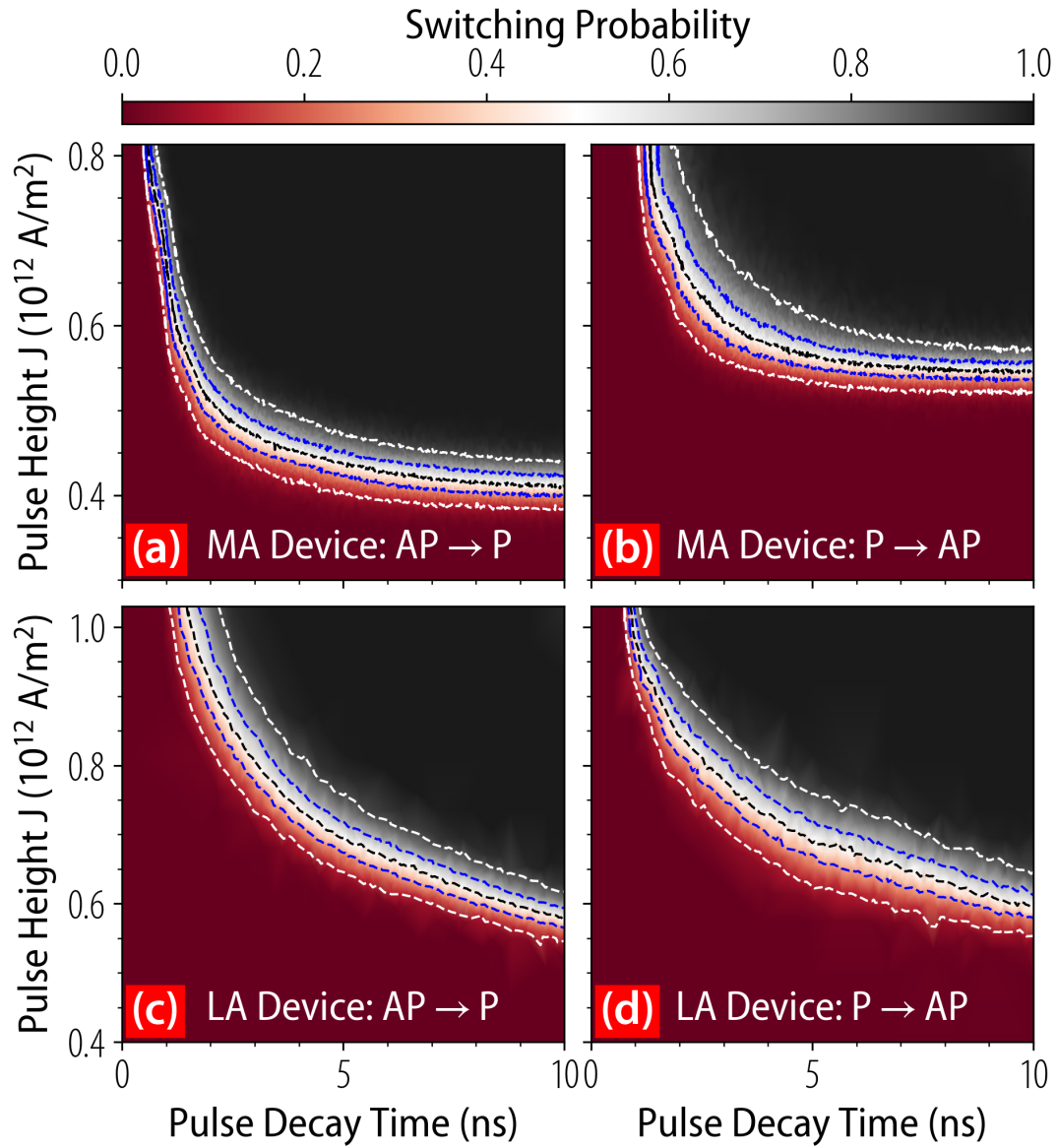

*Figure 3:* Phase diagrams for an (a,b) MA device and an (c,d) LA device switching in the (a,c) AP→P and (b,d) P→AP polarities. Color encodes the probability as shown at the top of the figure. The black dotted line shows the 50% probability boundary, while the blue and white dotted lines show ±20% and ±40% from the ±50% boundary, respectively. The image is composed of triangular elements drawn with a linear color-scale interpolation. Each vertex is the result of approximately 1024 switching attempts.

We measured the easy axis minor magnetic hysteresis loops of a standalone SHE-MTJ device using a superconducting coil pair, thereafter biasing the devices with an offset field that puts them in the bistable region supporting both parallel (P) and antiparallel (AP) resistance states. (This offset field was required since we have not yet optimized the SAF reference layer to reduce the fringing field seen by the FL to reliably be much less than its coercive field.) Two types of elliptical nanopillar samples were measured, those with a medium aspect (MA) ratio (nominally: 115nm x 45 nm) and those with a low aspect (LA) ratio (nominally: 115 nm x 70 nm). In Fig. 2, we show the minor loops for the field switching of the free layer of a MA and LA device. We observe sharp AP→P switching, but a curvature in the P→AP transition that we attribute to the Néel coupling between the two magnetic layers [19], and/or to the complex multidomain micromagnetic reversal process, as studied by simulation in [12], due at least in part to edge roughness.

We first determined the approximate thermal stability factors $\Delta(T)(\equiv E_a/k_B T)$ of the LA and MA devices by applying shallow pulses in the soft write-error regime [20], finding $\Delta_{\text{MA}}(3\text{K}) = 102 \pm 2$ and $\Delta_{\text{LA}}(3\text{K}) = 55 \pm 6$. (Here $E_a$ is the thermal activation energy for magnetic reversal.) We then performed short pulse switching attempts at an attempt rate of approximately 1 kHz, actively resetting the device with an inverted current pulse after each try.

Examples of the resultant switching phase diagrams (SPDs) are shown in Fig. 3(a-d) as a function of current density $J$ and pulse duration $\tau$. We used an adaptive measurement technique that iteratively refines a Delaunay triangulation [21] across the set of measurement coordinates, allowing us to acquire a very accurate phase diagram with many fewer points than are required for a rectilinear coordinate space. The SPDs demonstrate that for each polarity and pillar aspect ratio there is a region of highly successful switching (black). For the MA device there is a substantial difference in the current required to reach the 50% probability boundary (black dotted line) between AP→P and P→AP polarities, but much less so for the LA device. In all four cases the variation of the phase boundary, including its width $\delta J_{10-90}$, with decreasing pulse duration, does not exhibit the simple behavior that would be expected from rigid domain macrospin modeling; consistent with micromagnetic simulations [12] which show that in the strong, short-pulse regime the reversal typically proceeds by a distinctly non-uniform process involving the formation and propagation of sub-domains within the free layer. Those simulations showed the importance of the initial micromagnetic states, which differ substantially from the AP and P states. In the latter case (P→AP switching) if there was not a significant in-plane Oersted field the reversal was complex: requiring more current and exhibiting a broadened switching phase diagram. The AP→P reversal, on the other hand, was more coherent and proceeded more quickly. The application of a significant Oersted field in the simulations greatly sped up the P→AP reversal. Due to the high thermal stability of MA device at low temperature, the in-plane pulse field is not sufficiently large to impact the phase diagram as seen in Fig. 3(a-b). The LA device, meanwhile, shows little difference in $J_c$ across polarities, Fig. 3 (c-d), which we attribute to the greater effectiveness of the Oersted field in that case [12].

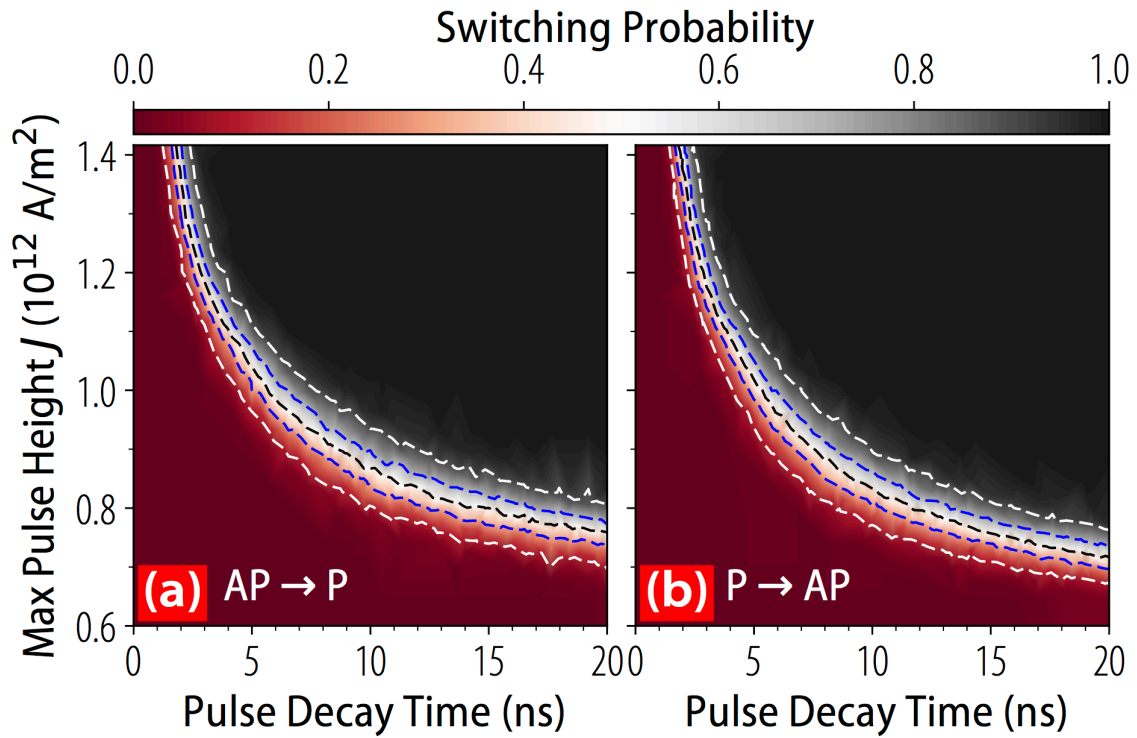

*Figure 4: Phase diagrams for a LA device subject to nTron-like pulses of (a) AP→P and (b) P→AP polarities. The left axes give the peak value of the current density J. The meaning of the dotted lines is described in Fig. 3.*

In the circuit of Fig. 1 the write current flows through the SHE-MTJ channel with an exponentially decaying profile in time. For traditional AD devices this presents complications: the initial energy from the write pulse is squandered given the collinearity of spin-torque and the lack of any thermally induced deflection of the FL from its energy minima. We expect better performance in the SHE-MTJ devices, however, given their Oersted-field assisted switching, and so we use a Keysight M8190A Arbitrary Waveform Generator (AWG) to stimulate our devices with exponentially decaying write pulses. We measure the switching phase-diagrams for our LA devices (Fig. 4), this time in terms of the peak current $J$ and the $1/e$ decay time of the pulses. We find that the switching phase boundary remains well defined, and observe an unexpected decrease in the polarity asymmetry observed in the rectangular phase diagrams (Fig. 3(a-b)). Both the positions and the profiles of the AP→P and P→AP phase boundaries converge: the P→AP profile becomes narrower at longer timescales and the AP→P profile becomes narrower at short timescales. Given the fine structure evident in our micromagnetic simulations, we conclude that STT during the exponential decay is most likely responsible for reliably coaxing the FL magnetization into its final state, thus reducing the width of the phase boundaries.

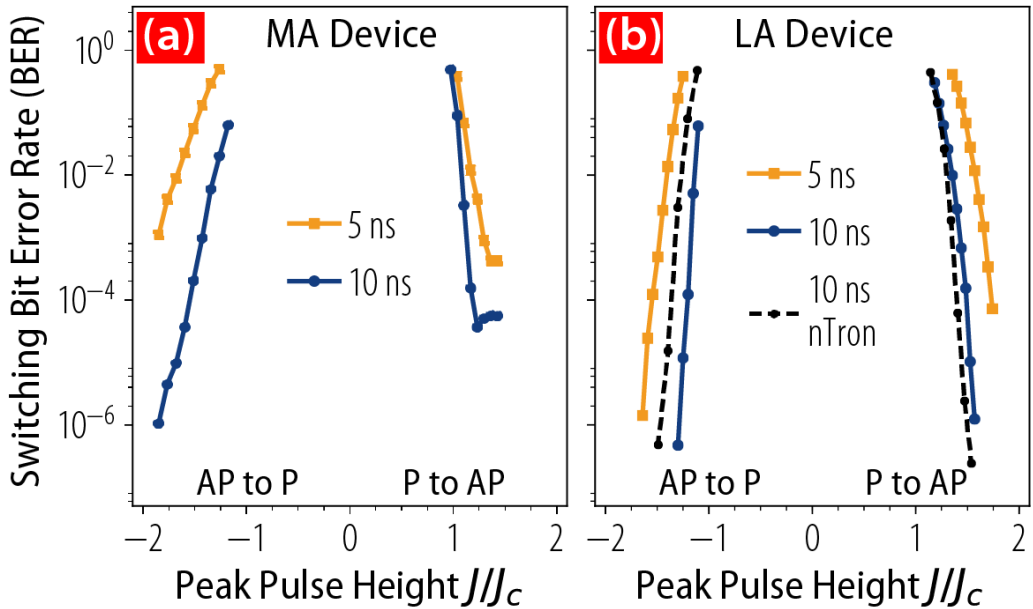

*Figure 5:* *Summary of the bit error rates (BERs) for LA devices. Peak pulse heights are normalized by the $J_c$ corresponding to their particular polarity and pulse envelope (rectangular or exponentially decaying).*

Finally, we examined the bit error rate (BER) of our devices at $T \approx 3\,\text{K}$. We applied an adaptive number of switching pulses that was sufficient to yield a switching probability at least one standard deviation above the statistical floor of our data (as determined by fitting to a $\beta$- distribution.) Once we found a BER of $10^{-6}$ the measurements were terminated — thus our BERs are bounded by measurement time constraints rather than any intrinsic limitations of the devices. The results for a LA device shown in Fig. 5 demonstrate a sharp falloff of the BER for both switching polarities; 10 ns rectangular pulses easily provide polarity-independent error rates of $< 10^{-6}$ for $J < 2J_c$. For 5 ns rectangular pulses, the LA devices start to exhibit some polarity dependence and suffer from a higher P→AP BER. The dotted line in Fig. 5 shows that the nTron-like pulses can produce similarly low error rates, again easily reaching our self-imposed limit of $< 10^{-6}$. This means that our intended memory cell (Fig. 1) is well suited to the particular stimulus requirements of the SHE-MTJs.

In closing, we have shown that SHE-MTJ devices can operate at cryogenic temperature with fast switching speeds ($\tau_0 \approx 1$ ns) and high reliability (BERs $< 10^{-6}$). As we previously observed at room temperature, this behavior is consistent with an assistive torque provided by the Oersted field [12]. It appears that an incubation period can still be avoided in the low temperature regime, owing to the non-collinearity of this additional initial torque. These results further confirm the advantages of Pt-based SHE-MTJ systems. We expect that further improvements in switching uniformity, speed and BER, could be obtained by utilizing devices whose $\Delta$ values are tuned to be $< 40$ at $T \approx 3$ K. We also found that our synthesized nTron-like switching pulses take particular advantage of this situation: the large initial amplitude of the pulses would be wasted were switching to proceed via the anti-damping spin torque alone. With these shaped pulses, we obtain polarity independent switching with BERs that taper off as rapidly as they do in response to rectangular pulses.

The research is based upon work supported by the Office of the Director of National Intelligence (ODNI), Intelligence Advanced Research Projects Activity (IARPA), via contract W911NF-14-C0089. The views and conclusions contained herein are those of the authors and should not be interpreted as necessarily representing the official policies or endorsements, either expressed or implied, of the ODNI, IARPA, or the U.S. Government. The U.S. Government is authorized to reproduce and distribute reprints for Governmental purposes notwithstanding any copyright annotation thereon. Additionally, this work was supported by the Office of Naval Research, by the NSF/MRSEC program (DMR-1120296) through the Cornell Center for

Materials Research, and by the NSF (Grant No. ECCS-0335765) through use of the Cornell Nanofabrication Facility/National Nanofabrication Infrastructure Network.

**Methods**

Samples are measured in a cryogen-free HPD cryo-probe station using microwave probes connected to lines thermalized by 0 dB attenuators at each thermal stage. We have confirmed that thermal noise from the ambient environment does not have an appreciate effect on our switching studies. Samples are stimulated using a combination of a PSPL 10,070A pulse generator with 2.5 ps time resolution and a Keysight M8190A 12GS arbitrary waveform generator. The former is passed through a voltage controlled attenuator giving us precise voltage control rather than relying on the coarse 1 dB step attenuators of the 10,070A. The two instruments' outputs are combined by a 180° hybrid before being routed to the cryostat.

Each switching attempt comprises a reset pulse of 10 ns, ~$1 \times 10^{12}$ A/m$^2$ in the opposite polarity to set the MTJ to its original state, followed by a switching pulse. The MTJ state is before and after each switching pulse with a small sense current of 3 μA across the junction. In switching phase diagrams shown in Figs. 3-4, the switching probability is given as the expectation value of the beta-distribution constructed from 1024 switching attempts. The bit error rates shown in Fig. 5 are determined in the same manner.

Our adaptive measurements are performed using our freely available Auspex and Adapt packages available at github.com/BBN-Q.